\documentclass{elsart}    

\usepackage{amsmath}
\usepackage{amssymb}
\usepackage{graphicx}
\usepackage{cite}
\usepackage{alltt}
\usepackage{bm} 
\usepackage{url}
\usepackage{longtable}
\usepackage{comment}


\begin{document}
\begin{frontmatter}

\title{{\tt GLISSANDO 3}:  {GL}auber {I}nitial-{S}tate {S}imulation {AND} m{O}re, ver.~3%
\thanksref{grant}} \thanks[grant]{Supported by Polish National Science Centre, 
grants 2015/17/B/ST2/00101 (PB), 2015/19/B/ST2/00937 (WB), and 2016/23/B/ST2/00692 (MR)}

\author[agh]{Piotr Bo\.zek},
\ead{Piotr.Bozek@fis.agh.edu.pl}
\author[as,ifj]{Wojciech Broniowski},
\ead{Wojciech.Broniowski@ujk.edu.pl}
\author[as]{Maciej Rybczy\'nski},
\ead{Maciej.Rybczynski@ujk.edu.pl}
\author[as]{Grzegorz Stefanek},
\ead{Grzegorz.Stefanek@ujk.edu.pl}

\address[agh]{AGH University of Science and Technology, Faculty of Physics and Applied Computer Science, 30-059 Cracow, Poland} 
\address[as]{Institute of Physics, Jan Kochanowski University, 25-406~Kielce, Poland} 
\address[ifj]{The H. Niewodnicza\'nski Institute of Nuclear Physics, Polish Academy of Sciences, 31-342 Cracow, Poland}

\begin{abstract}
We present ver.~3 of {\tt GLISSANDO}, a versatile Monte-Carlo generator for Glauber-like models of the initial 
stages of ultra-relativistic heavy-ion collisions. The present version incorporates the wounded parton model, 
which is phenomenologically successful in reproducing multiplicities of  particle production at the RHIC and the LHC. Within this model, one 
can study the nucleon substructure fluctuation effects, recently explored in p-A collisions. In addition, the code 
includes the possibility of investigating collisions of light nuclei, such as $^{3}$He and $^{3}$H, or  
the $\alpha$-clustered  $^{7,9}$Be, $^{12}$C, and  
$^{16}$O, where the deformation of the intrinsic wave function influences the transverse shape of the initial state. 
The current version, being down-compatible, 
retains the functionality of the previous releases, such as 
incorporation of various variants of Glauber-like models,
a smooth $NN$ inelasticity profile in the impact parameter obtained from a parametrization of experimental data, 
fluctuating strength of the entropy deposition, 
or realistic nuclear distributions of heavy nuclei with deformation. 
The code can provide output in the format containing the event-by-event source location, 
which may be further used in modeling the intermediate evolution phase, e.g.,  with hydrodynamics or transport models.
The interface is simplified, such that in the control input file the user may supply only the very basic information, 
such as the collision energy, the mass numbers of the colliding nuclei, and 
the model type. {\tt GLISSANDO 3} is integrated with the CERN {\tt ROOT} platform.
The package includes numerous illustrative and useful {\tt ROOT} scripts to compute and plot various results.
\end{abstract}

\date{12 January 2019, submitted to Comp. Phys. Comm.}


\end{frontmatter}

\maketitle

\newpage

{\bf PROGRAM SUMMARY}

\begin{small}
\noindent
{\em Program Title:  {\tt GLISSANDO~3} ver. 3.42}                                          \\
{\em Licensing provisions: CC By 4.0}                                  \\
{\em Programming language: {\tt C++} with the {\tt ROOT} libraries}                       \\


{\em Nature of problem:} 
The code implements in a versatile way the 
Glauber modeling of the initial stages of ultra-relativistic nuclear collisions, including the wounded nucleon and 
wounded quark models, with possible admixture of binary collisions. A state-of-the art inelastic nucleon-nucleon 
collision profile is implemented. A statistical distribution of the strength of the sources can be overlaid. 
The $\alpha$ clustered structure of light nuclei is built in.
\\
  
{\em Solution method:} 
Monte-Carlo simulation of nuclear collisions, analyzed off-line with numerous {\tt ROOT} scripts. The software allows for 
a straightforward event-by-event analysis of eccentricity coefficients and their correlations, size fluctuations, or multiplicity distributions. 
\\  
  
{\em Additional comments including Restrictions and Unusual features:} 
The input may consist only of the model type, the mass numbers of the nuclei and the collision energy.  
The output can also be used as initial conditions for further hydrodynamic studies.
\\


\end{small}  

\newpage 

\section{Introduction}

We present an extended version of the package 
{\tt GLISSANDO} -- {GL}auber {I}nitial-{S}tate {S}imulation {AND} m{O}re, previously described
in~\cite{Broniowski:2007nz} (ver.~1) and \cite{Rybczynski:2013yba} (ver.~2), where the user is directed for a more detailed 
description of the underlying physics and the features which remain unchanged in the present version.

With the on-going efforts to understand the nature of the ultra-relativistic nuclear collisions, more ideas and models are 
being explored. Consequently, the dedicated open-access software should follow these efforts and supply useful analysis tools. 
Concerning the earliest stage of such collisions, modeling based 
on the so-called Glauber approach~\cite{glauber1959high,Czyz:1969jg,Miller:2007ri} has been commonly used,
as it works well phenomenologically.

Another dedicated publicly available code implementing the Glauber approach is the PHOBOS Glauber
Monte Carlo~\cite{Alver:2008aq,Loizides:2014vua,Loizides:2017ack}, provided as a {\tt ROOT} package.
The Glauber initial condition is also an option in the parameterizations of {T\raisebox{-.5ex}{R}ENTo} code~\cite{Moreland:2014oya}.
Many other ultra-relativistic nuclear collision codes, 
e.g., HIJING~\cite{Wang:1991hta}, AMPT~\cite{Lin:2004en}, URQMD~\cite{Bass:1998ca}, or EPOS~\cite{Werner:2010aa} 
use internally the Glauber approach 
to model the early stage of the collision.

A very successful  variant of the Glauber model is based on 
{\em wounded quarks}~\cite{Bialas:1977en,Bialas:1977xp,Anisovich:1977av,Bialas:1978ze,Eremin:2003qn,KumarNetrakanti:2004ym,%
Adler:2013aqf,Adare:2015bua,Lacey:2016hqy,Bozek:2016kpf,Zheng:2016nxx,Mitchell:2016jio,Loizides:2016djv} acting
as the particle production sources, rather than the commonly 
used {\em wounded nucleons}~\cite{Bialas:1976ed,Bialas:2008zza} amended with {\em binary collisions}~\cite{Kharzeev:2000ph,Back:2001xy}). 
Implementation of the wounded quark (or, in general, {\em wounded parton}) model is the first major 
extension in {\tt GLISSANDO}~3.

Second, collisions of small nuclei with non-uniform distribution of nucleons may offer further stringent
tests of the geometry-flow  transmutation mechanism, including the event-by-event 
fluctuations~\cite{Ollitrault:1992bk,Aguiar:2001ac,Miller:2003kd,%
Manly:2005zy,Andrade:2006yh,Voloshin:2006gz,Alver:2006wh,Drescher:2006ca,%
Broniowski:2007ft,Voloshin:2007pc,Andrade:2008fa,Alver:2008zza,Broniowski:2009fm,Andrade:2009em,Hama:2009pk,Hirano:2009ah,Alver:2010gr,Alver:2010dn,%
Staig:2010pn,Teaney:2010vd,Qin:2010pf,Nagle:2010zk,Xu:2010du,Lacey:2010av,Werner:2010aa,Adare:2010ux,Qin:2011uw,Bhalerao:2011bp,Bhalerao:2011yg,%
Gardim:2011xv,Qiu:2011hf,Qiu:2011iv,Xu:2011jm,Teaney:2012ke,Jia:2012ma,Hirano:2012kj}. Moreover, 
the heavy-light reactions may provide insight into the 
ground-state clusterization structure in light nuclei, as first proposed in~\cite{Broniowski:2013dia,Bozek:2014cva} 
and further explored in~\cite{Zhang:2017xda,Rybczynski:2017nrx,Lim:2018huo}.  We thus offer a possibility of simulating reactions with 
$^{3}$He (studied experimentally in~\cite{Adare:2015ctn}) or $^{3}$H nuclei, which may be thought of as small triangles, as well as 
collisions involving light $\alpha$-clustered nuclei, such as  $^{7,9}$Be (dumbbell), $^{12}$C (triangle), and  $^{16}$O (tetrahedron). 

The nuclear configurations for $^{3}$He and $^{3}$H from~\cite{Wiringa:2013ala,Lonardoni:2017egu}, as supplied with the PHOBOS 
Glauber Monte Carlo input files~\cite{Loizides:2014vua}, are used in {\tt GLISSANDO}~3 for simulations with $A=3$ nuclei. 
Similarly, the user has an option to use the configurations for  $^{12}$C and  $^{16}$O
from~\cite{Wiringa:2013ala,Lonardoni:2017egu} as provided in~\cite{Loizides:2014vua}.

The user interface to the code has been greatly simplified. In  {\tt GLISSANDO}~3, it suffices to provide the type of the model used 
(wounded nucleons or wounded partons), the energy of the collision, and the mass numbers of the colliding nuclei. The appropriate
$NN$ inelastic cross section and its profile in the impact parameter are then automatically generated from  interpolation formulas based on the 
experimental $pp$ scattering data. Specifically, we use here the COMPETE model parametrization implemented in the Particle Data Group review
\cite{Patrignani:2016xqp}, which offers statistically the best description on the $pp$ and $p\bar p$ scattering data over 
a very broad range of the collision energies.  

As in the previous releases, the output of {\tt GLISSANDO}~3 is generated in two ways. The key features of the formed initial states
are stored in a {\tt ROOT} tree containing ready to use results in a compact form. In addition, a full event-by-event output 
can be generated for use in event-by-event studies of further evolution stages, such as hydrodynamics~\cite{Andrade:2006yh,Werner:2009fa,Petersen:2010cw,%
Holopainen:2010gz,Gardim:2011xv,Bozek:2011if,Schenke:2010rr,Qiu:2011fi,Chaudhuri:2011pa}
or transport.

The new features implemented in {\tt GLISSANDO 3} include: 

\begin{itemize}
 \item State-of-the art nucleon-nucleon inelastic cross sections and their impact-parameter profiles.
 \item Implementation of the wounded quark (in general, wounded parton) model.
 \item Possibility of colliding $^{3}$He and $^{3}$H, with the distributions from external files.
 \item Inclusion of $\alpha$-clustered structure of $^{7,9}$Be, $^{12}$C, and  $^{16}$O nuclei.
 \item Possibility of studying the effect of the proton fluctuations in the framework of the 
          wounded parton model.
 \item Simplified user interface.
\end{itemize}

The main retained features of the previous releases incorporate:

\begin{itemize}
 \item Parametrization of distributions of all popular nuclei, in particular those used in the energy - system-size scan
      of the NA61/SHINE~\cite{na61url:2013}.
 \item Inclusion of the nuclear deformation~\cite{Filip:2007tj,Filip:2009zz,Filip:2010zz}. 
 \item Possibility of using externally-generated (correlated) nuclear distributions, e.g.,~\cite{Alvioli:2009ab,Broniowski:2010jd} for $^{208}$Pb, 
  $^{197}$Au or $^{40}$Ca.
 \item Possibility of studying proton-nucleus and deuteron-nucleus collisions.
 \item Possibility of overlaying weights over the distribution of sources (Poisson, Gamma, negative binomial).
 \item Study of the core-corona effect~\cite{Hohne:2006ks,Becattini:2008ya,Bozek:2005eu,Werner:2007bf}.
 \item Output of the event-by-event data with location of the sources to a text file, to be used for initialization of hydrodynamics of transport codes.
 \item A reference manual generated with {\tt doxygen}~\cite{doxygen:2013} may be useful for those who wish to tailor the code.
\end{itemize}

\section{New features in {\tt GLISSANDO 3}}

\subsection{Inelastic cross sections and inelasticity profiles \label{sec:inel}}

With  new measurements of the total, elastic, and differential elastic cross section, in particular at 
the LHC~\cite{Antchev:2013paa,Antchev:2015zza,Aad:2014dca} and in cosmic rays~\cite{Collaboration:2012wt}, 
new parameterizations of the $pp$ and $p\bar{p}$ scattering amplitudes became available. 
A standard description is provided by the COMPAS group within the COMPETE model, presented in 
Sec.~52 of the 2016 edition of the Review of Particle Physics~\cite{Patrignani:2016xqp}, which parametrizes the $pp$ and $p\bar{p}$ 
scattering data in a very accurate way. 
In {\tt GLISSANDO 3} we use this parameterization of the data as the most complete one.

\begin{figure}[tb]
\begin{center}
\includegraphics[width=1\textwidth]{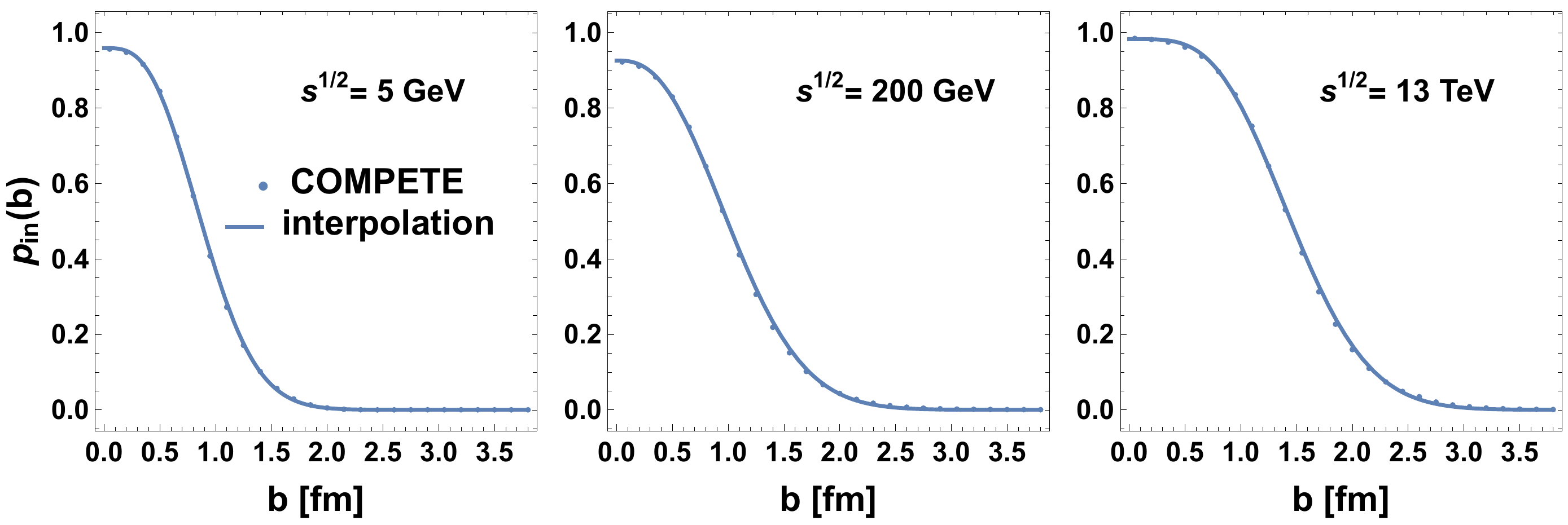}
\end{center}
\vspace{0mm}
\caption{
The inelastic profile plotted as a function of the impact parameter for three sample $pp$ collision  energies $\sqrt{s}$. 
The points indicate the COMPETE model parametrization, whereas the lines show the interpolation according to Eq.~(\ref{eq:inter}).
\label{fig:inter}}
\end{figure}

The inelastic profile function, 
determining the probability of an inelastic $NN$ collision at impact parameter $b$ and collision energy $\sqrt{s}$ is defined as 
\begin{eqnarray}
p_{\rm in} (b,s)  = \sigma_{\rm tot} (b,s)-\sigma_{\rm el} (b,s)= 4p_{\rm CM} \, {\rm Im}\,  h(b,s) -  4p_{\rm CM}^2|h(b,s)|^2,   \label{eq:prof}
\end{eqnarray} 
where $h(b,s)$ is the Fourier-Bessel transform of the $pp$ elastic scattering amplitude $f(s,-q^2)$, 
\begin{eqnarray}
2p_{\rm CM} h(b,s)=  2\int_0^\infty \!\!\! q dq J_0(bq) f(s,-q^2), \label{eq:invf}
\end{eqnarray} 
with $p_{\rm CM}=\sqrt{s/4-m_N^2}$ denoting the CM momentum of the nucleon. Using the COMPETE model for 
$f(s,-q^2)$ with parametrization from~\cite{Patrignani:2016xqp}, we 
thus obtain $\sigma_{\rm in} (b,s)$. In the code, rather than using these long formulas involving a numerical integration in 
Eq.~(\ref{eq:invf}), we apply a simple parametrization of the inelastic profile of the form~\cite{Rybczynski:2013mla},
\begin{eqnarray}
p_{\rm in}(s,b) =  
G \Gamma\left (\frac{1}{\omega(s)}, \frac{\pi G(s) b^2}{\omega(s) \sigma_{\rm in}(s)} \right ) /
\Gamma \left(\frac{1}{\omega(s)}\right), \label{eq:Gamma}
\end{eqnarray}
where $\Gamma(a,z)$ denotes the incomplete Euler $\Gamma$ function, $\sigma_{\rm in}(s)$ is the inelastic cross 
section, and $G(s)$ and $\omega(s)$ are suitably adjusted parameters such that the 
COMPETE results are accurately reproduced. The values obtained from our fit can be efficiently represented with interpolating 
functions (which we use for $\sqrt{s}\ge 5$~GeV):
\begin{eqnarray}
\sigma_{\rm in}(s) &=& [40.32 (\sqrt{s}/{\rm GeV}+ 53.08)^{0.104}-30.15-8.75/(\sqrt{s}/{\rm GeV})]{\rm ~mb}, \label{eq:inter} \\
G(s) &=& \frac{33.82+1.27 (\sqrt{s}/{\rm GeV})^{0.85}}
   {32.10  (\sqrt{s}/{\rm GeV})^{0.063}+1.28  (\sqrt{s}/{\rm GeV})^{0.85}-10^{-8}}, \nonumber  \\
\omega(s) &=& \frac{-3.97+4.41 (\sqrt{s}/{\rm GeV})^{0.24}}
   {1+4.41  (\sqrt{s}/{\rm GeV})^{0.40}-0.27  (\sqrt{s}/{\rm GeV})^{0.24}}. \nonumber
\end{eqnarray}
Note that the applied expressions are for shear numerical convenience and do not reflect physical mechanisms of high-energy 
$pp$ scattering.

\begin{figure}[tb]
\begin{center}
\includegraphics[width=0.46\textwidth]{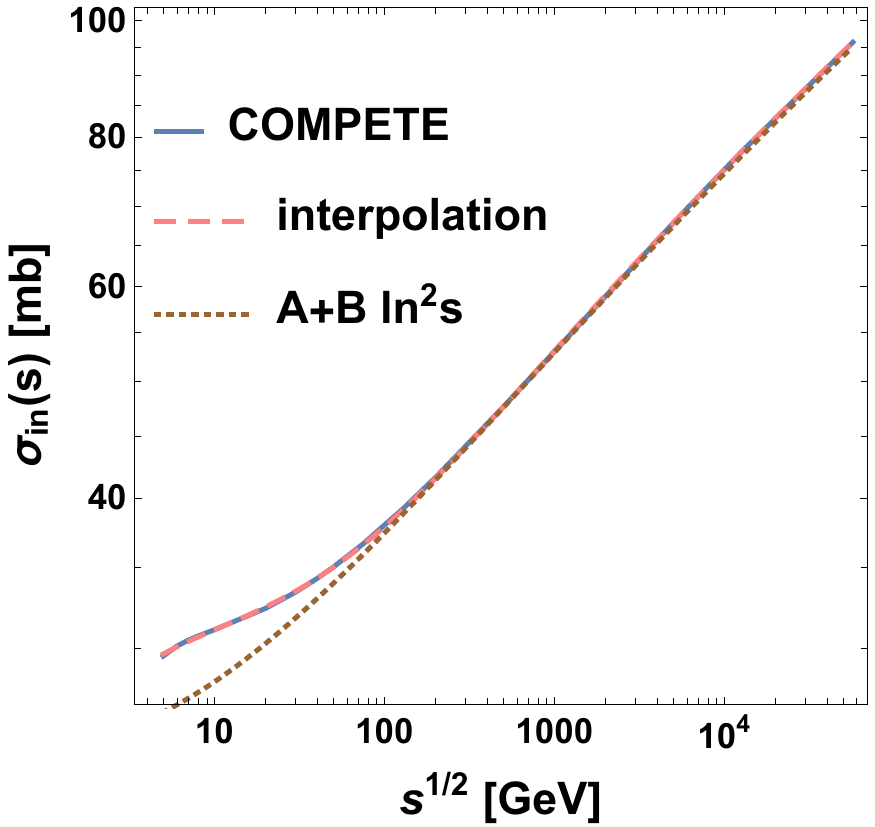} \hspace{.7cm}  \includegraphics[width=0.46\textwidth]{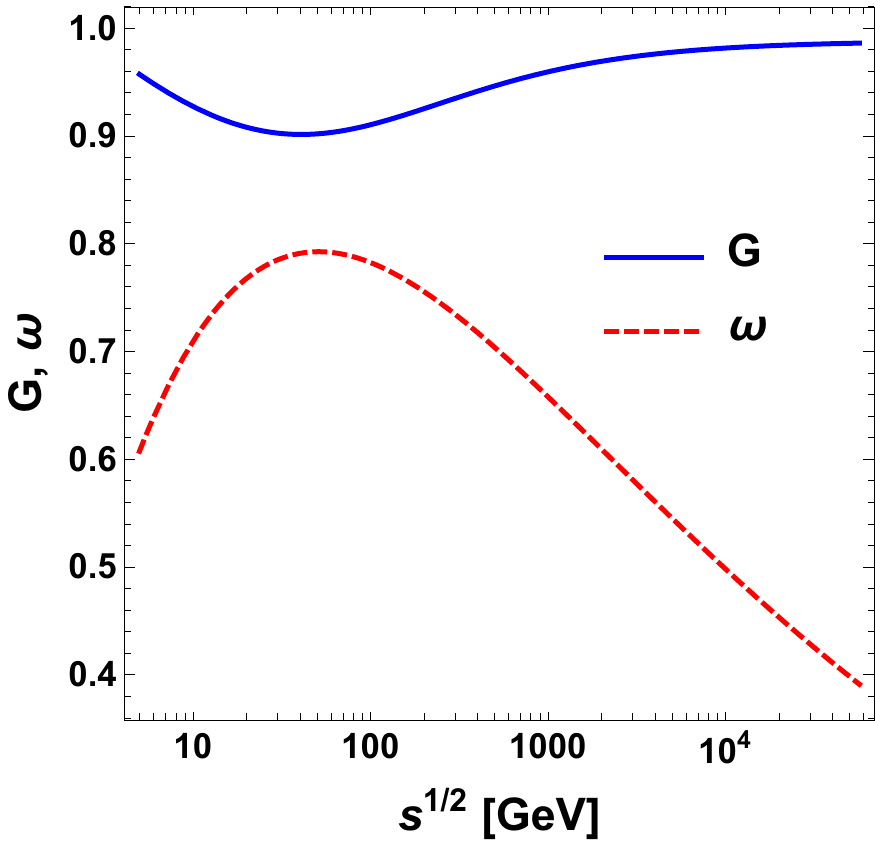} 
\end{center}
\vspace{0mm}
\caption{
The inelastic $pp$ cross section (left) and parameters $G$ and $\omega$ (right), 
plotted as functions of the collision energy. See the text for discussion. 
\label{fig:parNN}}
\end{figure}

The quality of the interpolating formulas is illustrated in Fig.~\ref{fig:inter}, where we compare the inelastic profile functions 
from the COMPETE model to Eq.~(\ref{eq:Gamma}) with parameters (\ref{eq:inter}) for three selected collision energies.
In the left panel of Fig.~\ref{fig:parNN} we show the results of the COMPETE model for the total inelastic $pp$ cross section $\sigma_{\rm in}(s)$ 
(solid line) and our simple fit of Eq.~(\ref{eq:inter}). We also show (dotted line) the parametrization 
$A+B\, {\rm ln}^2(s/{\rm GeV}^2)$, with $A=25$~mb and $B=0.146$~mb, used in~\cite{Loizides:2017ack}. We note that whereas at higher collision 
energies ($\sqrt{s}>100$~GeV) this parametrization agrees with the COMPETE model (and thus with the data) accurately, for 
$\sqrt{s}<100$~GeV the COMPETE parametrization, or our Eq.~(\ref{eq:inter}), is clearly advantageous.
The interpolated dependence of parameters $G$ and $\omega$ on the collision energy, as given by formulas  (\ref{eq:inter}), 
is displayed in the right panel Fig.~\ref{fig:parNN}.

We recall that the use of a realistic $NN$ inelasticity profile (as opposed to sometimes used hard-sphere profile 
in older software, e.g.,~\cite{Wang:1991hta}) is phenomenologically 
important. First, it leads to proper elastic differential cross section in $pp$ collisions, as shown above. Second, 
it yields sizable effects in standard observables in ultra-relativistic nuclear collisions. In particular, 
the effect on ellipticity of the fireball was discussed in detail in~\cite{Rybczynski:2011wv}, where 
the use of a realistic (smooth) wounding profile led to a significant (10-20\%) reduction for peripheral A-A collisions, compared to the 
case with a hard-sphere profile.

\subsection{Wounded partons \label{sec:quarks}}

In the wounded quark model~\cite{Bialas:1977en,Bialas:1977xp,Bialas:1978ze,Anisovich:1977av} the 
valence quarks play the role of elementary scatterers. The approach is implemented in {\tt GLISSANDO 3}, as 
it has proved to be phenomenologically successful in describing the multiplicity of produced hadrons in a variety of reactions,
with the linear scaling 
\begin{eqnarray}
\frac{dN_{\rm ch}}{d\eta} \sim Q_{\rm W}
\end{eqnarray}
well satisfied. Here $N_{\rm ch}$ denotes the observed charged hadrons, and $Q_{\rm W}$ the number of wounded quarks 
in a given reaction and centrality class. 
It was first noticed that the RHIC data follow the scaling~\cite{Eremin:2003qn}, further
explored at RHIC by the PHENIX Collaboration~\cite{Adler:2013aqf,Adare:2015bua}, as well as 
at the SPS~\cite{KumarNetrakanti:2004ym}. 
See also~\cite{Loizides:2014vua,Adler:2013aqf,Adare:2015bua,Bozek:2016kpf,Lacey:2016hqy,Zheng:2016nxx,Mitchell:2016jio,Loizides:2016djv}
for further developments. 

Our implementation of the model is described in detail in~\cite{Bozek:2016kpf}. The nucleon consists of $p$ partons 
($p=3$ for the wounded quark model) distributed according to a radial density
\begin{eqnarray}
\rho(r)={\rm const} \; r^2 \exp \left [ -\sqrt{\tfrac{2}{3} \left (1-\tfrac{1}{p}\right)}\,\frac{r}{r_0(s;p)}\right].  \label{eq:eps}
\end{eqnarray}
The scale parameter $r_0(s;p)$ controls the size of the nucleon built of partons. 
The factor $1-1/p$ accounts for the center-of-mass 
corrections; a shift to the center of mass of the nucleon is carried out after randomly generating the positions of the centers of the partons.
As $r_0$ is adjusted  phenomenologically (see the following), the presence of the $1-\tfrac{1}{p}$ factor in formula (\ref{eq:eps}) is conventional. 

In the wounded parton approach one needs to choose the parton-parton inelasticity profile. We take it in the Gaussian
form
\begin{eqnarray}
p_{\rm in}^{qq}(s,b) = \exp \left[{- \frac{\pi b^2}{\sigma_{\rm in}^{qq}(s;p)}}\right], \label{eq:Gauss}
\end{eqnarray}
where $\sigma_{\rm in}^{qq}(s)$ is the parton-parton inelastic cross section. 

\begin{figure}[tb]
\begin{center}
\includegraphics[width=0.85\textwidth]{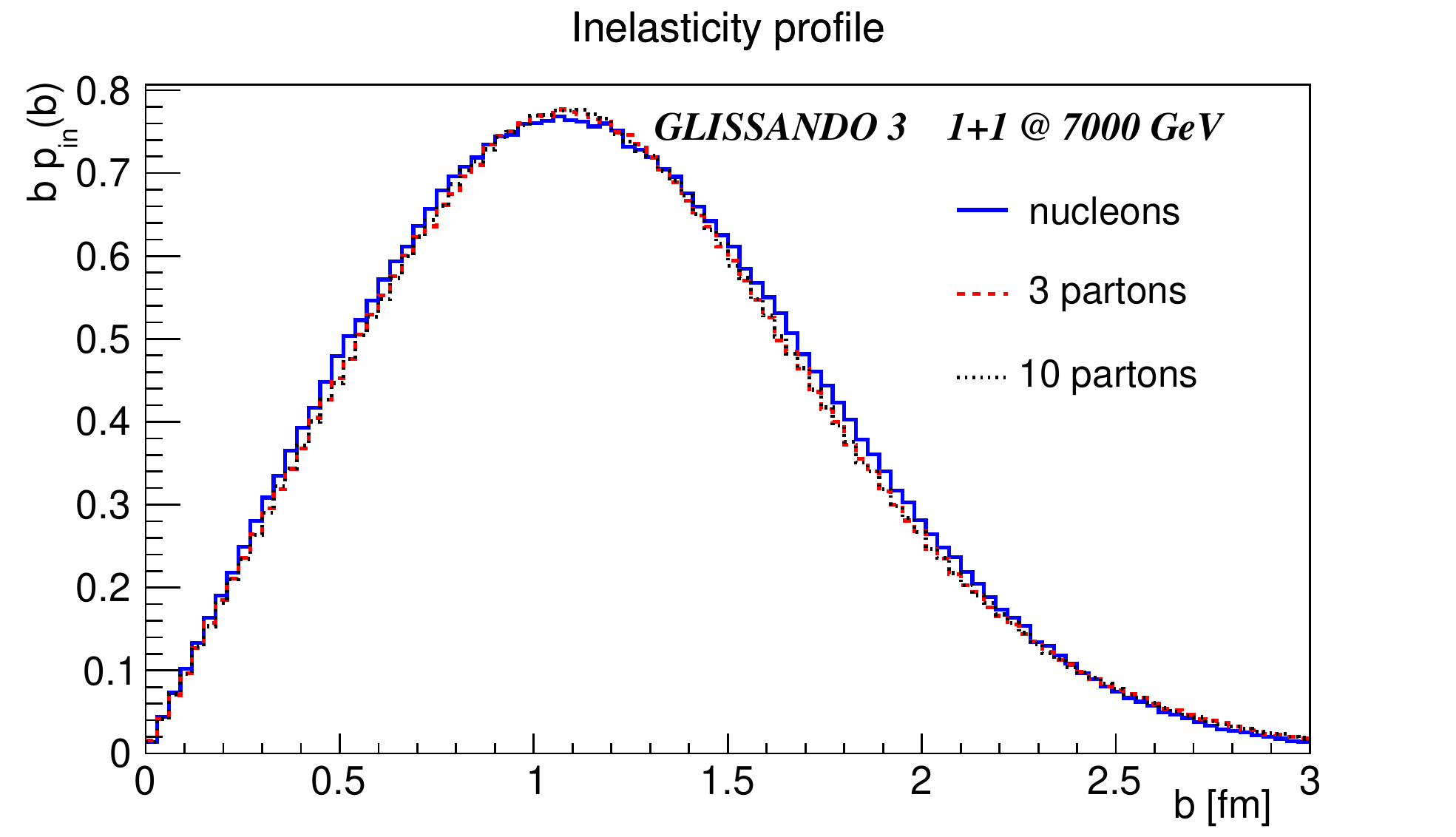}
\end{center}
\vspace{-5mm}
\caption{
The inelastic $NN$ profiles (multiplied with $b$) obtained in the wounded nucleon and the wounded parton model for $p=3$ and $p=10$
(generated with script {\tt inel\_prof.C}). 
\label{fig:NNqq}}
\end{figure}

Thus our wounded parton model implementation brings in two parameters:  $r_0(s;p)$ and $\sigma_{\rm in}^{qq}(s;p)$. Their 
values are chosen in such a way that the resulting $NN$ collision profile reproduces the COMPETE parametrization 
discussed in Sec.~\ref{sec:inel}. In other words, we impose the desired feature that the $NN$ inelasticity profile is the same  
if one uses nucleons or partons as elementary scatterers. With our chosen parameterizations of Eq.~(\ref{eq:eps}) and 
(\ref{eq:Gauss}) this cannot be accomplished in an exact manner, but optimization leads to a
close agreement (within a few percent). This can be seen in Fig.~\ref{fig:NNqq}, where we compare the $NN$ elasticity profiles 
(multiplied with $b$)
obtained in the nucleon and parton models for $\sqrt{s}=7$~TeV. The quality of agreement is similar at other 
collision energies or values of $p$.

\begin{figure}[tb]
\begin{center}
\includegraphics[width=0.46\textwidth]{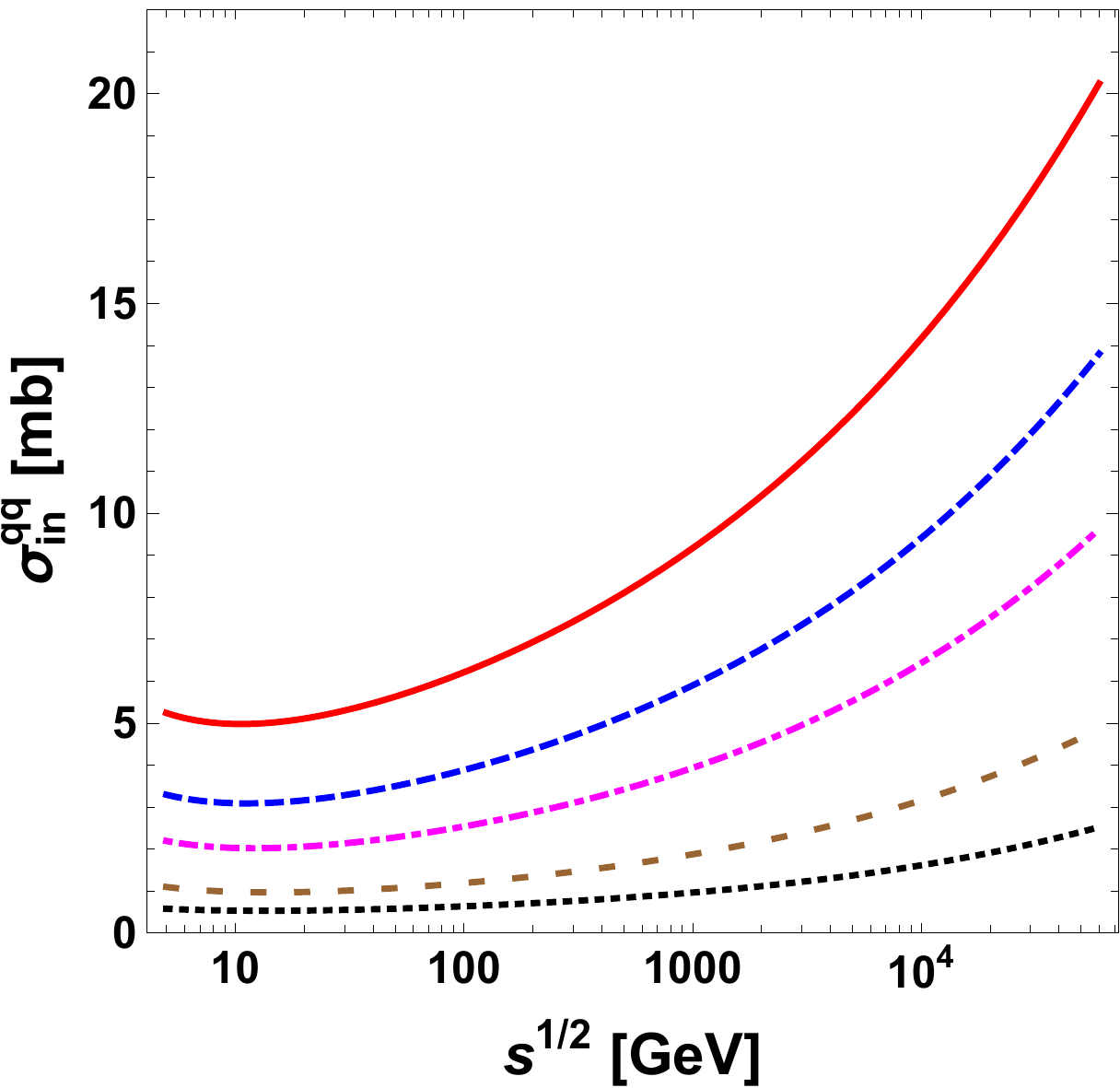} \hspace{.7cm} \includegraphics[width=0.46\textwidth]{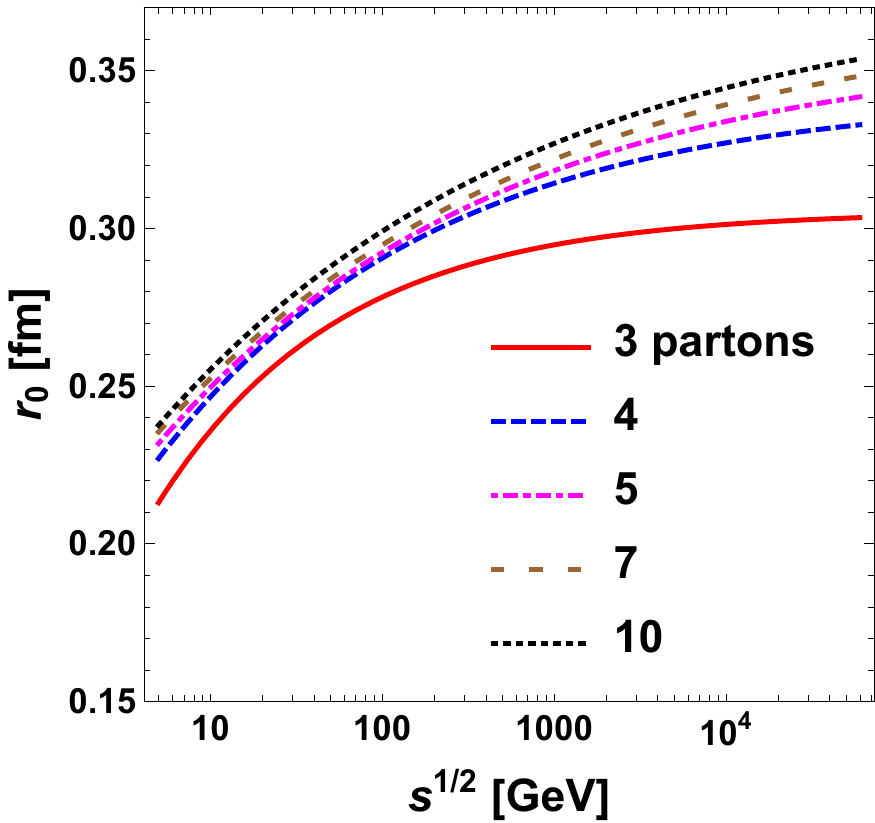}
\end{center}
\vspace{0mm}
\caption{
The dependence of the inelastic parton-parton cross section (left) and the parton size distribution parameter $r_0$ (right) on the collision energy.
\label{fig:qs}}
\end{figure}

The optimum values of the wounded parton model parameters are interpolated as follows for the parton-parton 
cross section and the parton distribution parameter:
\begin{eqnarray}
&& \sigma_{\rm in}^{qq}(s;p) =[A_1 + A_2 (\sqrt{s}/{\rm GeV})^{A_3} + A_4/(\sqrt{s}/{\rm GeV})]{\rm ~mb}, \label{eq:interq1}\\
&& ~~~~A_1=-1.66 + 8.73/p + 0.11 p, \;\; A_2=-1.58 + 9.03/p + 0.08 p, \nonumber \\
&& ~~~~A_3=0.22 - 0.07/p + 0.01 p, \;\; A_4=-1.02 + 23.10/p - 0.05 p, \nonumber 
\end{eqnarray}
\begin{eqnarray}
&& r_0(s;p) = [B_1 + B_2 (\sqrt{s}/{\rm GeV})^{B_3}]{\rm ~fm}, \label{eq:interq2}\\
&& ~~~~B_1=0.37 - 0.95/p^2 + 0.11, \;\; B_2=-0.21 - 0.39/p^2 + 0.22/p, \nonumber \\ 
&& ~~~~B_3=-0.28 - 5.21/p^2 + 1.36p. \nonumber
\end{eqnarray}
The above parameterizations work for $3 \le p \le 10$.
The dependence of $\sigma_{\rm in}^{qq}(s;p)$ and $r_0(s;p)$ on $\sqrt{s}$ is visualized in Fig.~\ref{fig:qs}.
We note that in our model the parton-parton cross section decreases with $p$, whereas the dependence of the 
size parameter $r_0$ exhibits a rather weak growth with $p$. We also note that an approximate scaling $\sigma_{\rm in}^{qq}(s;p) \sim 1/p^2$ holds, 
which is accurate to a few percent for $p>3$. 

As in previous releases of our code, 
centers of nucleons in a nucleus are distributed according to a Woods-Saxon density with additional $NN$ 
repulsion~\cite{Broniowski:2007nz}, or are taken from external calculations, {\em e.g.},~\cite{Alvioli:2009ab} of~\cite{Loizides:2014vua}.
Nuclear shape deformation~\cite{Heinz:2004ir,Filip:2007tj,Filip:2009zz,Rybczynski:2012av} is implemented
for deformed nuclei, such as $^{63}$Cu, $^{197}$Au, or $^{238}$U, in the same manner as in {\tt GLISSANDO 2}.

\subsection{Reactions with light clustered nuclei \label{sec:alpha}}

\begin{figure}[b]
\begin{center}
\includegraphics[width=0.5\textwidth]{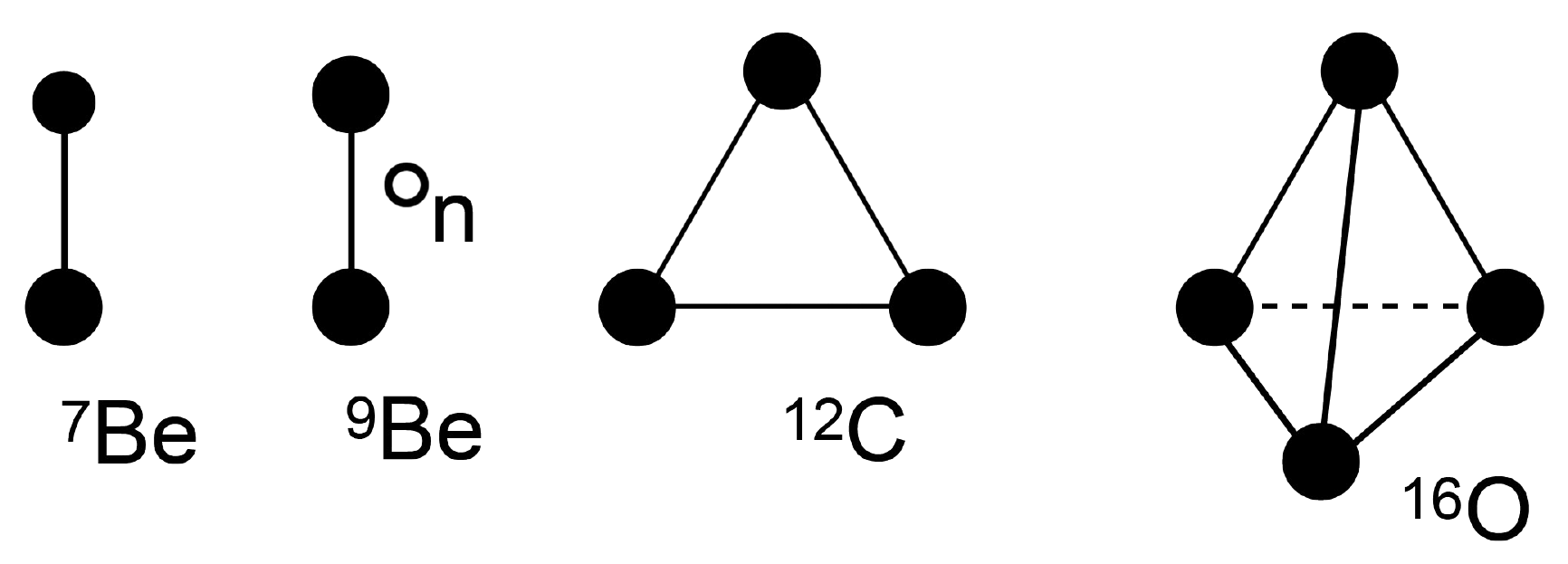}
\end{center}
\vspace{0mm}
\caption{
Cartoon of the geometry of light clustered nuclei (from~\cite{Rybczynski:2017nrx}). The bigger filled blobs represent the $\alpha$ cluster, the smaller 
filled blob the ${}^3$He cluster, and the empty blob the extra neutron in ${}^9$Be. 
\label{fig:clus}}
\end{figure}

{\tt GLISSANDO 3} offers a possibility to collide nuclei whose structure exhibits clustering. 
Specifically, we implement $^{7}$Be as $\alpha +^{3}$He, $^{9}$Be as $2\alpha+$neutron, 
$^{12}$C as $3\alpha$, and $^{16}$O as $4\alpha$ (see Fig.~\ref{fig:clus}). The details of 
our fixing of geometric parameters of the distributions are presented in~\cite{Rybczynski:2017nrx}.
The parameters are adjusted in such way that the one-body nuclear distributions~\cite{Wiringa:2013ala,Lonardoni:2017egu} are matched. 
We first arrange the positions of clusters as in Fig.~\ref{fig:clus}, separating the centers
by the distance $l$. The distribution of the centers of nucleons in each cluster is randomly generated according to a Gaussian
\begin{eqnarray}
f_i(\vec{r})={\rm const} \; \exp \left (- \frac{3}{2} \, \frac{(\vec{r}-\vec{c_i})^2}{r_c^2} \right ),   
\end{eqnarray}
where $\vec{r}$ denotes the coordinate of the nucleon and
$\vec{c_i}$ is the position of the center of the cluster $i$. The width of the cluster is controlled with $r_c$, the rms radius 
of the cluster. Then the positions of the nucleons are generated sequentially, switching between 
clusters 1, 2,\dots, 1, 2,\dots, until all the needed nucleons are placed.

\begin{table}[tb]
\caption{\label{tab:paral} Parameters of the distribution of light clustered nuclei used in {\tt GLISSANDO 3}.}
\vspace{3mm}
\begin{center}
\begin{tabular}{|c|cccc|}
\hline
 Nucleus & $l$ [fm] & $r_\alpha$ [fm] &  $r_{{}^3{\rm He}}$ [fm]  & $r_n$ [fm]
\\
\hline 
$^{7}$Be   & 3.2 & 1.2 & 1.4 & - \\
$^{9}$Be   & 3.6 & 1.1 & -  & 1.9  \\
$^{12}$C   & 2.8 & 1.1 & - & - \\
$^{16}$O   & 3.2 & 1.1 & - & - \\
\hline
\end{tabular}
\end{center}
\end{table}

For the case of ${}^{9}$Be we place the extra neutron on top of the two $\alpha$ clusters. 
The distribution is 
\begin{eqnarray}
f_n(\vec{r})={\rm const} \;  r^2 \,\exp \left (- \frac{3}{2} \frac{r^2}{r_n^2} \right ),
\end{eqnarray}
and it exhibits a hole in the middle.

The values of parameters used in the code for the cluster model are collected in Table~\ref{tab:paral}.

\begin{figure}[tb]
\begin{center}
\includegraphics[width=0.65\textwidth]{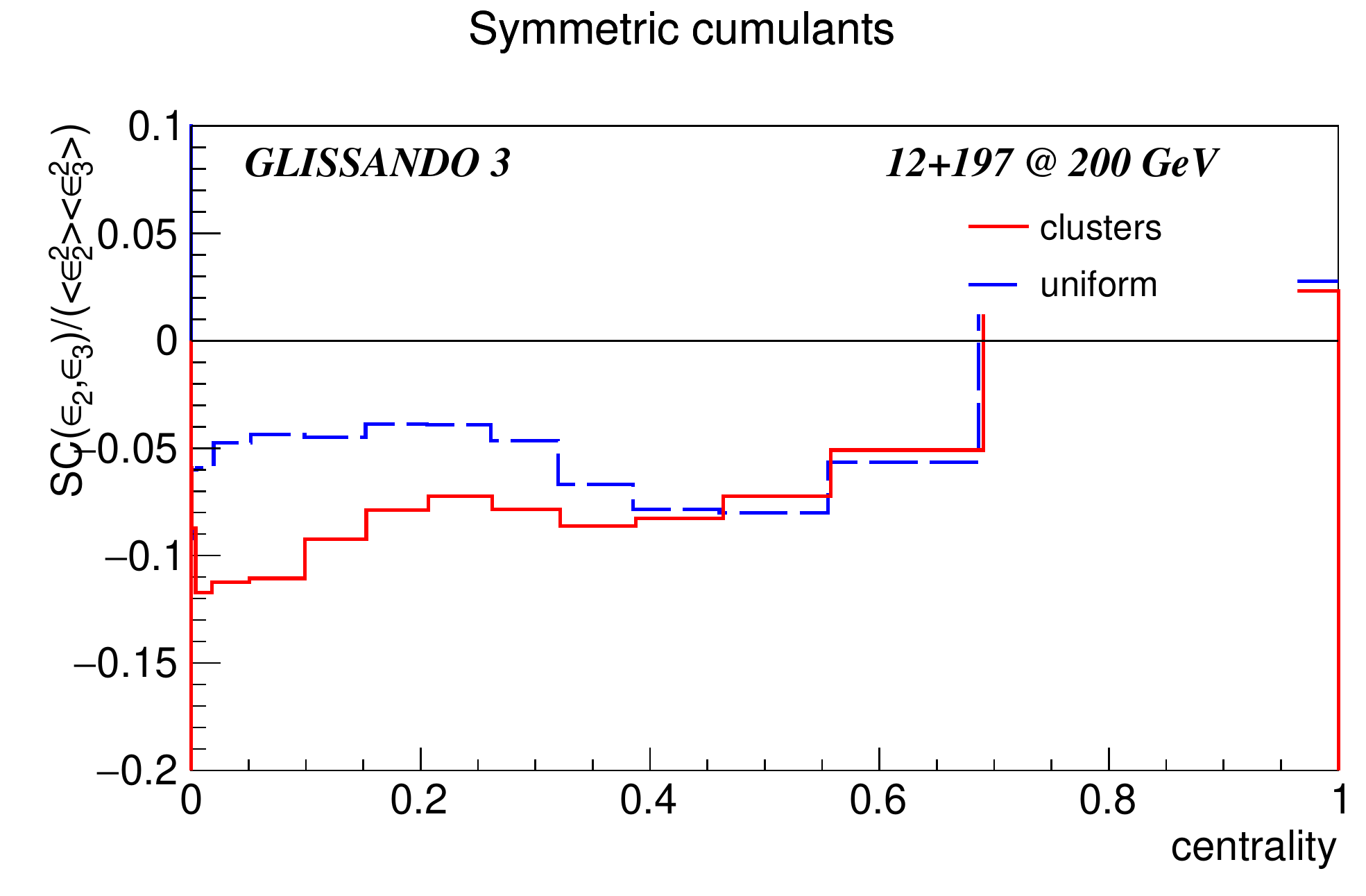}
\end{center}
\vspace{0mm}
\caption{
The scaled symmetric cumulant of eccentricities in collisions of clustered and uniform $^{12}$C nucleus with 
$^{197}$Au. The symmetric cumulant is defined as $SC(a,b)=\langle a^2b^2\rangle -\langle a^2\rangle \langle b^2\rangle$  
(generated with script {\tt sc\_comp.C}).
\label{fig:sc}}
\end{figure}

One may do many physical studies with light clustered nuclei, in particular looking for clusterization signatures 
in harmonic flow patterns~\cite{Broniowski:2013dia,Bozek:2014cva,Rybczynski:2017nrx,Zhang:2018zzu}. 
In Fig.~\ref{fig:sc} we present an example of a quantity, the symmetric cumulant of 
elliptic and triangular flows, which displays different behavior on the model with clusters
compared to the case without clusters (uniform). 

\vspace*{0.7cm}

\subsection{Smearing of sources \label{sec:smear}}

The Monte Carlo Glauber model provides, event-by event, the positions of point-like sources (wounded objects) in the transverse plane.
Physically, the sources must have transverse size, which is accomplished by {\em smoothing} them, typically with a Gaussian form
\begin{eqnarray} 
\Delta(x,y;x_i,y_i)= \frac{1}{\pi \sigma^2}\exp \left[-\frac{(x-x_i)^2+(y-y_i)^2}{\sigma^2} \right ],  \label{eq:gau} 
\end{eqnarray}
where $(x,y)$ are the transverse coordinates and  $(x_i,y_i)$ gives the location of the $i$th source, and $\sigma$ is the smearing width.
Smearing is necessary, {\em i.a.},  in forming the initial condition for subsequent hydrodynamic evolution.    
Moreover, smearing increases the size of the fireball and reduces the eccentricities, which are its 
basic characteristics. The effect is particularly relevant for small systems. 
The effects of Gaussian smearing (\ref{eq:gau}) are simple to implement, as the corresponding 
expressions are analytic.

The eccentricity of rank $n$ ($n\ge2$) in a given event, $\epsilon_n$, and the corresponding angle of the principal 
axis (the event-plane angle), $\Psi_n$, are defined in the standard way as
\begin{eqnarray}
\epsilon_n e^{i \Psi_n} \equiv - \frac{\int dx \, dy \, s(x,y) \rho^n e^{i n \phi}}{\int dx \, dy\, s(x,y) \rho^n }, \label{eq:epsigen}
\end{eqnarray}
where $s(x,y)$ is the transverse entropy distribution, and we have introduced the polar coordinates via $x=\rho \cos \phi$, $y=\rho \sin \phi$.
When the sources are point-like, then $s(x,y)=\sum_i \delta(x-x_i)\delta(u-y_i)$ and
\begin{eqnarray}
\epsilon_n e^{i \Psi_n} = - \frac{\sum_i \rho_i^n e^{i n \phi_i}}{\sum_i \rho_i^n }, \label{eq:epsi_point}
\end{eqnarray}
where $i$ runs over all the sources in the event, and $x_i=\rho_i \cos \phi_i$, $y_i=\rho_i \sin \phi_i$.
With Gaussian smearing (\ref{eq:gau}), we have $s(x,y)=\sum_i \Delta(x,y;x_i,y_i)$ and, correspondingly, 
\begin{eqnarray}
&& \epsilon_n e^{i \Psi_n} =  - \frac{\sum_i N_i^{(n)}}{\sum_i D_i^{(n)}}, \\
&& N_i^{(n)}= \int dx \, dy \, \Delta(x,y;x_i,y_i) \rho^n e^{i n \phi}, \;\; D_i^{(n)}= \int dx \, dy \, \Delta(x,y;x_i,y_i) \rho^n. \nonumber \label{eq:epsi_sm}
\end{eqnarray}

In polar coordinates we may write explicitly
\begin{eqnarray}
N_i^{(n)}=\frac{1}{\pi \sigma} \int\!\! \rho \, d\rho \int\!\! d\phi \exp \left(-\frac{\rho^2+\rho_i^2 -2 \rho \rho_i \cos(\phi - \phi_i)}
{\sigma^2} \right) \rho^n e^{i n (\phi-\phi_i)} e^{i n \phi_i}. \nonumber \\
\end{eqnarray}
Integration over $\phi$ yields 
\begin{eqnarray}
N_i^{(n)}= \frac{2}{\sigma ^2} \int\!\! \rho \, d\rho \, { \rho^{n+1} e^{-\frac{\rho ^2+\rho _i^2}{\sigma ^2}} I_n\left(\frac{2 \rho  \rho_i}{\sigma ^2}\right)} \,  e^{i n \phi_i}, 
\end{eqnarray}
and integration over $\rho$ 
\begin{eqnarray}
N_i^{(n)}=   \rho_i^n  e^{i n \phi_i}, 
\end{eqnarray}
which means that the numerator with smearing in Eq.~(\ref{eq:epsi_sm}) is exactly the same as for point-like sources in Eq.~(\ref{eq:epsi_point}).

Repeating the above steps for the denominator yields
\begin{eqnarray}
&& \!\!\!\!\!  D_i^{(n)}=\frac{1}{\pi \sigma} \int\!\! \rho \, d\rho \int\!\! d\phi \exp \left(-\frac{\rho^2+\rho_i^2 -2 \rho \rho_i \cos(\phi - \phi_i)}{\sigma^2} \right) \rho^n  \\
&& = \frac{2}{\sigma ^2} \int\!\! \rho \, d\rho \, { \rho^{n+1} e^{-\frac{\rho ^2+\rho _i^2}{\sigma ^2}} I_0\left(\frac{2 \rho  \rho_i}{\sigma ^2}\right)} = 
\sigma ^n \Gamma \left(\frac{n}{2}+1\right) L_{\frac{n}{2}}\left(-\frac{\rho_i^2}{\sigma ^2}\right), \nonumber
\end{eqnarray}
where $\Gamma(z)$ is the Euler Gamma function and $L_a(z)$ is the Laguerre polynomial. The forms of $D_i^{(n)}$ for the first few even values of $n$ are
\begin{eqnarray}
&& D_i^{(2)} = \rho_i^2+\sigma ^2, \\
&&  D_i^{(4)} =\rho ^4+4 \rho_i^2 \sigma ^2+2 \sigma^4, \nonumber \\
&&  D_i^{(6)} = \rho_i^6+9 \rho_i^4 \sigma ^2+18 \rho_i^2 \sigma ^4+6 \sigma ^6, \nonumber \\ && \dots \nonumber 
\end{eqnarray}
For odd values of $n$
\begin{eqnarray}
&& D_i^{(3)} = \frac{1}{4\sigma} \sqrt{\pi } e^{-\frac{\rho_i^2}{2 \sigma ^2}} \left[2 \rho_i^2 \left(\rho_i^2+
2 \sigma ^2\right) I_1\left(\frac{\rho_i^2}{2 \sigma ^2}\right) \right .  \nonumber \\ && \hspace{3.5cm} \left . +\left(2 \rho_i^4+6 \rho ^2 \sigma 
   ^2+3 \sigma ^4\right) I_0\left(\frac{\rho_i^2}{2 \sigma ^2}\right)\right], \nonumber \\
&&   D_i^{(5)} = \frac{1}{8\sigma} \sqrt{\pi } e^{-\frac{\rho_i^2}{2 \sigma ^2}} \left[\rho_i^2 \left(4 \rho_i^4+
24 \rho_i^2 \sigma ^2+23 \sigma ^4\right) I_1\left(\frac{\rho_i^2}{2 \sigma ^2}\right) \right . \nonumber \\ && \hspace{3.5cm} \left . +\left(4 \rho_i
   ^6+28 \rho_i^4 \sigma ^2+45 \rho_i^2 \sigma ^4+15 \sigma ^6\right) I_0\left(\frac{\rho_i^2}{2 \sigma ^2}\right)\right], \label{eq:bes} \\ && \dots \nonumber 
\end{eqnarray}
where $I_{0,1}(z)$ are the modified Bessel functions of the first kind. 
It is apparent from the above formulas  
that smearing increases the denominators, therefore $\epsilon_n$ is quenched, in agreement with intuition. At the same time we observe that the principal axes angles $\Psi_n$ 
are not altered by smearing.
\begin{figure}[tb]
\begin{center}
\includegraphics[width=0.5\textwidth]{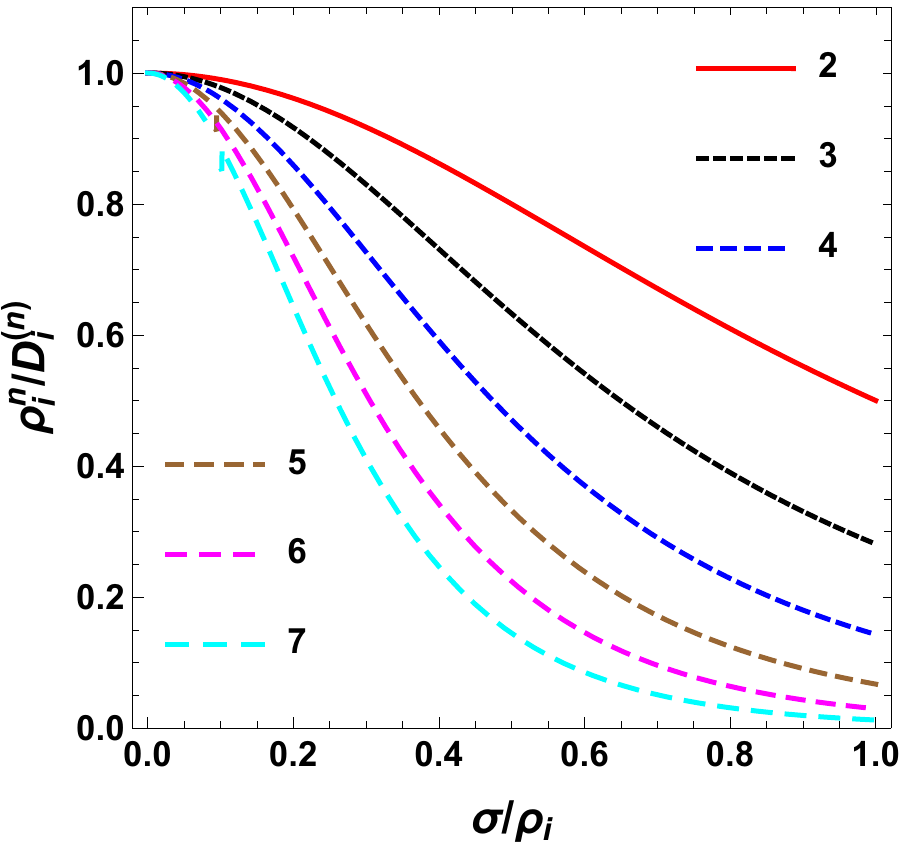}
\end{center}
\vspace{0mm}
\caption{The reduction of the contribution to eccentricities $\epsilon_n$ from a source placed at $\rho_i$, plotted versus $\sigma/\rho_i$ for $n=2,3, \dots, 7$.
\label{fig:redu}}
\end{figure}
\vspace*{0.7cm}

The reduction of the contribution of a single source placed at $\rho_i$ for first few eccentricities, quantified with $\rho_i^2/D_i^{(n)}$, is shown in Fig.~\ref{fig:redu}. 
We note that, as expected, the effect increases with $n$ and with the value of the smearing parameter $\sigma$.

Taking for definiteness the ellipticity $(n=2)$ and collecting the contributions from all sources, we find the 
smearing reduction factor $\langle r^2 \rangle /(\langle r^2 \rangle +\sigma^2)$, which becomes increasingly important with decreasing mean squared radius, $\langle r^2 \rangle$, 
of the fireball. 

Smearing is incorporated in {\tt GLISSANDO 3} in the evaluation of eccentricities and the mean squared radius of the fireball. The amount of smearing is controlled with the parameter 
{\tt DS}$=\sigma/2$, whereby it has the interpretation of the Gaussian width in one dimension.

\subsection{Rapidity modeling \label{sec:rap}}

The modeling in rapidity, present to some extent in ver.~2 of the code, has been removed. This is
because, unlike the modeling in the transverse plane where the Glauber approach became one 
of the standards, the longitudinal modeling has much more freedom and is under development. If the user 
wishes to model the rapidity dependence of the fireball, she/he may use the transverse distributions
(generated event-by-event and written to an output file) in an external application.

\section{Installation and running \label{sec:install}}

The user must first install the CERN {\tt ROOT} package~\cite{root1}.
After downloading and unpacking {\tt GLISSANDO 3}, the command

\begin{verbatim}
make
\end{verbatim}

\noindent should be run, which creates the executable binary  {\tt glissando3}. 

To optionally recreate the {\tt doxygen} manual (provided as {\tt /doc/latex/refman.pdf} with 
the distribution), the {\tt doxygen}~\cite{doxygen:2013} package should be installed and  the following commands executed: 

\begin{verbatim}
make cleandoc
make doc
\end{verbatim}

With the code installed, the user may look at an instructive presentation of the main features of the present version
by executing the shell script

\begin{verbatim}
./demo_ver_3.sh
\end{verbatim}

\noindent To see the created {\tt pdf} files {\tt evince} should be installed prior to the run. 
To use another pdf viewer, the line {\tt VIEW=evince} should be edited in the {\tt *.sh} files.
The features of ver.~2, retained in the present release, may be explored by executing 

\begin{verbatim}
./demo_ver_2.sh
\end{verbatim}

For collisions of two nuclei, the standard running command has the syntax

\begin{verbatim}
./glissando3 [input_file] [output_file]
\end{verbatim}

\noindent where the nuclei can be any, including the proton and the deuteron.
When the input or output arguments are absent, the defaults are 
\begin{verbatim}
input.dat - default input
glissando.root - default output
\end{verbatim}
The input parameters with their defaults are collected in Appendix~\ref{sec:input}.

\subsection{Makefile}

The {\tt Makefile} contains commands for compilation and linking. It can be used for profiling the code by modifying the 
preprocessor options. Most importantly, the user should decide if she/he uses the wounded nucleon or the wounded parton version of the model.  
In the nucleon case the standard preprocessor options should read

{\small
\begin{verbatim}
PREPROCESS   := -D_partons_=0 -D_nnwp_=2 
  -D_bindep_=1 -D_files_=0 -D_profile_=0 -D_weight_=0 -D_evout_=0 
  -D_clusters_=0 -D_uncluster_=0 -D_pardis_=0 -D_rdsconv_=1
\end{verbatim}
}

\noindent whereas for the wounded parton model

{\small
\begin{verbatim}
PREPROCESS   := -D_partons_=1 -D_nnwp_=1 
  -D_bindep_=1 -D_files_=0 -D_profile_=0 -D_weight_=0 -D_evout_=0 
  -D_clusters_=0 -D_uncluster_=0 -D_pardis_=0 -D_rdsconv_=1
\end{verbatim}
}

The meaning of all preprocessor parameters is as follows: 
{\small
\begin{verbatim}
 _nnwp_=0 - hard-sphere NN wounding profile, =1 - Gaussian, =2 - gamma
 _bindep_=1 - binary and wounding depend on each other, =0 - do not
 _files_=1 - read the nuclear distributions from external files, 
           =0 - generate randomly
 _profile_=1 - generate the nucleon profile and NN correlation data, =0 - do not
 _weight_=1 - generate data for the NN collision profiles 
                      and for the weight (RDS) ditributions, 0 - do not 
 _evout_=0 - do not generate the text event data, =1 - short, =2 -long
 _clusters_=1  - generate clusters in light nuclei, =0 do not
 _uncluster_=1 - smoothe out the alpha clusters in nucleus A, 
                   =2 - in nucleus B, 
                   =3 in both nuclei, 
                   =0 - do not smoothe
 _partons_=1 - wounded parton model, =0 - wounded nucleon model
 _pardis_=0  - exponential parton distribution in nucleus, =1 - Gaussian
 _rdsconv_=2 - convention for RDS: 1 - Nw/2, 2 - Nw
\end{verbatim}
}

The user may modify the {\tt Makefile} file to suit his needs, or, alternatively, run {\tt make} for example as follows:
\begin{verbatim}
make 'PREPROCESS = -D_partons_=1 -D_nnwp_=1 -D_evout_=1 '
\end{verbatim}
to produce the binary code for the wounded parton model with the Gaussian wounding profile and with generation of some results
to an external ASCII file. 

Another functionality is the storage of the current version of the package,
\begin{verbatim}
make package
\end{verbatim}
as well as the cleaning options:
\begin{verbatim}
make clean
make cleandoc
make cleanoutput
\end{verbatim}

\subsection{Input}

The input file is a standard ASCII file. Every line contains the name of the parameter separated with space from the 
assigned value. The sequence of the lines with parameters is flexible.  
When a parameter is missing in the input file, or a line containing it is commented out with \#, the default 
value is used. See Appendix~\ref{sec:input} for details concerning the input parameters.

\subsection{Output}

\noindent 
A typical output from a run of {\tt GLISSANDO 3} to the console is shown in Table~\ref{tab:out}.

\begin{table}[tb]
\caption{A typical output to the console from {\tt GLISSANDO3}. \label{tab:out}}
\begin{center}
{\footnotesize
\begin{verbatim}

user@host: ~/GL3$  ./glissando3 
Start: Mon Nov 12 15:36:16 2018
--------------------------------------

*********************************************************
GLISSANDO 3 ver. 3.3
ver. 3: http://arxiv.org.abs/1901.04484
ver. 2: Computer Physics Communications 185 (2014) 1759, arXiv:1310.5475
ver. 1: Computer Physics Communications 180 (2009)   69, arXiv:0710.5731
**********************************************************
Simulation of nuclear collisions in Glauber models
----------------------------------------------------------
parameters reset from default in input/input.dat:
EVENTS	2000
NUMA	208
NUMB	208
ECM	2760

generates ROOT output file output/glissando.root

number of events: 2000

208+208 @ sqrt(s_NN)=2760GeV

Woods-Saxon: RA=6.40677fm, aA=0.459fm, dA=0.9fm 
Woods-Saxon: RB=6.40677fm, aB=0.459fm, dB=0.9fm 

NUCLEON MODEL
mixed model: sig_w=61.9917mb, sig_bin=61.9917mb, alpha=0.12
RDS distribution scale with u=1 (wounded) and u=1 (binary)
gamma wounding profile, G=0.972569, omega=0.586854
source smearing width: 0.4 fm
acceptance window: b_min=0fm, b_max=25fm, Nw_min=2, Nw_max=100000

event: 2000     (100%)              

Some quantities for the specified acceptance window 
(+/- gives the e-by-e standard deviation):
A+B cross section = 7226.7mb  (makes sense for min. bias)
efficiency (accepted/all) = 36.8053%
N_w = 113.994+/-114.84
RDS: 93.5426+/-107.285

Finish: Mon Nov 12 15:37:04 2018
(0h:0m:48s)
**************************************

\end{verbatim}
}
\end{center}
\end{table}

The results of the simulation are stored in the {\tt ROOT} output file or, optionally, also in text files when the preprocessor parameter 
{\tt\_evout\_}=1 or 2 is used.  

To see the physical results, the user should enter 
the {\tt ROOT} environment
\begin{verbatim}
root
\end{verbatim}
\noindent
and execute one of the supplied scripts {\tt *.C}
\begin{verbatim}
.x macro/demo_3/<script name>.C(<optional parameters>)
\end{verbatim}
\noindent
or directly examine the output {\tt ROOT} file, e.g. with a 
\begin{verbatim}
TBrowser a 
\end{verbatim}
\noindent
command (executed within {\tt ROOT} environment).
An alternative method of executing the {\tt ROOT} scripts is provided in the example shell {\tt demo\_ver\_3.sh}.
Proceeding this way does not require a greater familiarity with {\tt ROOT}.

The preprocessor option {\tt \_evout\_} controls the possible 
event-by-event output to an additional external ASCII file with the name of the {\tt ROOT} 
output file appended with {\tt .points}. If {\tt \_evout\_=2}, then the file stores the transverse positions 
event-by-event. Its content consists of the blocks 

\begin{verbatim}
RDS   NwAB   specA   atan2(ycmA,xcmA)   specB   atan2(ycmB,xcmB)
\end{verbatim}
followed with {\tt NwAB} lines of the format
\begin{verbatim}
x  y  z  c  w 
\end{verbatim}
The second part of the block is
\begin{verbatim}
Nbin  
\end{verbatim}
followed with {\tt Nbin} lines of the format
\begin{verbatim}
x  y  z  c  w 
\end{verbatim}

\noindent Above we use the notation {\tt RDS} for the relative deposited strength in the event (defined below), 
{\tt NwAB} for the number of wounded objects (nucleons or partons), {\tt specA} for the number of spectators from nucleus {\tt A} and   
{\tt atan2(ycmA,xcmA)} for the azimuthal angle of their center of mass relative to the $x$ axis, 
and {\tt specB} and {\tt atan2(ycmB,xcmB)} for analogous quantities for nucleus {\tt B},
and {\tt Nbin} for the number of binary collisions.
Next, {\tt x} and {\tt y} are the transverse coordinates of the sources 
in fm, {\tt z} is the longitudinal coordinate, {\tt |c|} indicates how many times a wounded nucleon collided, with 
positive (negative) {\tt c} corresponding to nucleus {\tt A} ({\tt B}), while {\tt c=0} indicates the binary collisions. The last entry is the weight {\tt w}
of the given source (by {\em source} we customarily mean a wounded nucleon or a binary collision).
The number of blocks is equals to the number of events.

If {\tt \_evout\_=1}, then a simplified output is generated, with the format 

\begin{verbatim}
NwA   NwB   Nbin   RDS 
\end{verbatim}
\noindent where  {\tt NwA} and  {\tt NwB} are the numbers of wounded objects in nuclei A and B, respectively.
The number of output lines equals to the number of events.

The {\em relative deposited strength} (term introduced in~\cite{Broniowski:2007nz}) is equal to 
\begin{eqnarray}
{\tt RDS} =  \frac{1-\alpha}{2} {\tt NwAB} + \alpha \, {\tt Nbin}, \label{eq:rds} 
\end{eqnarray}
in accordance to the Glauber model with the binary collision admixture~\cite{Kharzeev:2000ph,Back:2001xy}.
Correspondingly, the weight of a given source is $(1-\alpha)/{2}$ for the wounded objects and  $ \alpha$ for the binary collisions. 

\subsection{Reading external nuclear distributions}

The user may provide external files with nuclear distributions.
We have used two sets of such files. The first one comes from 
Alvioli et al.~\cite{Alvioli:2009ab} for the case of  $^{16}$O, $^{40}$Ca, and $^{208}$Pb. They can be obtained from 
{\tt http://sites.psu.edu/color/}. The user must create from the downloaded files a singe file, for instance running 

\begin{verbatim}
cat ca40-1.dat ca40-2.dat ca40-3.dat [more files] > ca40.dat
\end{verbatim}

\noindent The resulting file {\tt o16.dat} should be placed in the relative subdirectory {\tt nucl}.

The files have the format

\begin{verbatim}
x   y   z   k
\end{verbatim}

\noindent where x, y, and z denote the Cartesian coordinates of the centers on nucleons in fm, while $k=0$ for neutrons and $k=1$ for 
protons. Groups of {\tt A} lines, where {\tt A} is the mass number of the nucleus, correspond to a single nuclear configuration.

The second case concerns $^3$He or $^3$H distributions, which are taken from 
the Green's function Monte Carlo simulations~\cite{Carlson:1997qn} as provided in~\cite{Loizides:2014vua}. For the 
user's convenience the files {\tt he3\_plaintext.dat} and {\tt h3\_plaintext.dat} are also provided at the {\tt GLISSANDO} web site.
The format of these files is 

\begin{verbatim}
x1   y1   z1   x2   y2   z2   x3   y3   z3   foo  foo  foo  foo 
\end{verbatim}

\noindent where {\tt (xi, yi, zi)} are the Cartesian coordinates of the $i$th nucleon in a given configuration.  

Similarly, the configurations of $^{12}$C or $^{16}$O from~\cite{Carlson:1997qn} as provided in~\cite{Loizides:2014vua} may be 
used. They are provided at the {\tt GLISSANDO} web site as {\tt carbon\_plaintext.dat} and {\tt oxygen\_plaintext.dat}.

To use the nuclear configurations from external files, the code must be compiled with

\begin{verbatim}
make 'PREPROCESS = -D_files_=1'
\end{verbatim}

\noindent and executed as 

{\small
\begin{verbatim}
./glissando3 [input_file] [output_file] [nucleus_A_file] [nucleus_B_file] 
\end{verbatim}
}

\noindent When the syntax 

\begin{verbatim}
./glissando3 [input_file] [output_file] [nucleus_A_file]
\end{verbatim}

\noindent is used, then the distribution of the nucleons in nucleus {\tt A} is read from an external file, 
whereas for nucleus {\tt B} the positions of nucleons are generated randomly. This syntax must
also be used for the collisions of nucleus A with the proton or the deuteron.

\subsection{Fixing centrality cuts \label{sec:cencut}}

We explain how to impose centrality cuts on  {\tt GLISSANDO 3} simulations 
and how to properly choose the range for the impact parameter $b$ (a too large range slows down the simulations). The supplied shell script {\tt centrality.sh}
implements the following procedure. 
First, a minimum-bias simulation must be run, with no (or broad-range) values for the {\tt W0}, {\tt W1}, 
{\tt RDS0}, and {\tt RDS1} parameters, as well as {\tt BMIN=0} and {\tt BMAX} set to 
a larger value than the sum of the radii of the two colliding nuclei. We typically use {\tt BMAX} as about twice
this sum to account for the tails in the Woods-Saxon distributions. Note, however, that large values for {\tt BMAX}
lead to a larger Monte Carlo rejection rate,slowing down the simulation.
Next, the {\tt macro/demo\_3/cent.C} script should be executed in root. 
The values of {\tt W0}, {\tt W1}, or {\tt RDS0}, {\tt RDS1} corresponding to the desired centrality classes can be read off from the 
generated file {\tt output/centrality.dat}. Next, the input file 
must be modified with the proper values for {\tt W0}, {\tt W1} supplied (if centrality 
is determined by the multiplicity of the wounded objects), or {\tt RDS0}, {\tt RDS1} (if centrality 
is given by the relative deposited strength RDS), and the code must be rerun (with low statistics). 
Running {\tt /macro/demo\_3/b\_dist.C} and 
looking at the plot  {\tt b\_dist.pdf} allows us to determine the optimum values for {\tt BMIN=0} and {\tt BMAX}.
We thus should update these parameters in the input file and rerun simulations with full statistics. 

The above steps can be simply traced by executing {\tt centrality.sh}.

\section{Structure of the code}

The structure of the code is provided in the supplied reference manual created with {\tt doxygen}. 

\section{Summary}

We have presented {\tt GLISSANDO 3}, with hopes it will continue to be a useful and versatile tool for the 
heavy-ion community. The authors most welcome suggestions and questions from the users.

We would like to thank dr Maksym Deliyergiyev for his help to make our scripts compatible with ROOT6.
\bigskip

This research was carried out in laboratories created under the project
"Development of research base of specialized laboratories of public 
universities in Swietokrzyskie region"
no POIG 02.2.00-26-023/08 dated May 19, 2009.

\appendix

\section{Input and output \label{sec:input}}

The input parameters are collected in Table~\ref{tab:input}. The hash sign \# at the 
beginning of the line in the input file comments it out, in which case a default value of the parameter as set in the code 
(file {\tt functions.h}) is used.

{\small

\begin{longtable}{lrp{4in}}
\caption{Parameters in the input file. \label{tab:input}}\\
\hline
{name} & 
{default} & 
{description} \\
\hline \vspace{-3mm}
\endfirsthead

\multicolumn{3}{c}{{\tablename} \thetable{} -- Continued} \\[0.5ex]
\hline \hline
{name} & 
{default} & 
{description} \\
\hline \vspace{-3mm}
\endhead


\multicolumn{3}{c}{{Continued on Next Page\ldots}} \\
\endfoot


\hline \hline
\endlastfoot


{ECM}    & {-1}    & {center of mass energy of the colliding system [GeV]} \\\vspace{-3mm}
{NUMA}   & {208}   &  {mass number of nucleus $A$} \\\vspace{-3mm}
{NUMB}   & {208}   &  {mass number of nucleus $B$} \\\vspace{-3mm}
{EVENTS} & {10000} &  {number of generated events} \\\vspace{-3mm}
{BMIN}   & {0.}    &  {minimum impact parameter [fm]}\\\vspace{-3mm}
{BMAX}   & {25.}   &  {maximum impact parameter [fm]}\\\vspace{-3mm}
{W0}     & {2}     &  {minimum allowed number of wounded nucleons}\\\vspace{-3mm}
{W1}     & {100000}  &  {maximum allowed number of wounded nucleons}\\\vspace{-3mm}
{RDS0}   & {0}     &  {minimum allowed RDS}\\\vspace{-3mm}
{RDS1}   & {100000} &  {maximum allowed RDS}\\\vspace{-3mm}
{MODEL}  & {0}     &  {0 - constant superimposed weight, 1 - Poisson, 2 - Gamma, 3 - Negative Binomial} \\\vspace{-3mm} 
{Uw}     & {1. }   &  {Poisson, Gamma, or Negative Binomial mean value for wounded objects}\\\vspace{-3mm}
{Ubin}   & {1.}     &  {Poisson, Gamma, or Negative Binomial mean value for binary collisions}\\\vspace{-3mm}
{Vw}     & {2.}     &  {Gamma or Negative Binomial variance for wounded objects}\\\vspace{-3mm} 
{Vbin}   & {2.}     &  {Gamma or Negative Binomial variance for binary collisions}\\\vspace{-3mm} 
{NCS}    & {3}     &  {number of partons in the nucleon} \\\vspace{-3mm}
{NBIN}   & {40}    &  {number of bins for histograms in $\rho$, $x$, or $y$} \\\vspace{-3mm}
{RWSA}   & {-1} &  {Woods-Saxon radius for the distribution of centers, nucleus $A$ [fm]} \\\vspace{-3mm}
{AWSA}   & {0.459} &  {Woods-Saxon width, nucleus $A$ [fm]}\\\vspace{-3mm}
{BETA2A} & {-1}   &  {deformation parameter $\beta_{2}$, nucleus $A$}\\\vspace{-3mm}
{BETA4A} & {-1}   & {deformation parameter $\beta_{4}$, nucleus $A$}\\\vspace{-3mm}
{ROTA\_THETA} & {-1} & {rotation parameter (angle $\theta$), -1 - random rotation, nucleus $A$} \\\vspace{-3mm} 
{ROTA\_PHI} & {-1} & {rotation parameter (angle $\phi$), -1 - random rotation, nucleus $A$} \\\vspace{-3mm} 
{RWSB}   & {-1} &  {Woods-Saxon radius for the distribution of centers, nucleus $B$ [fm]}\\\vspace{-3mm}
{AWSB}   & {0.459} &  {Woods-Saxon width, nucleus $B$ [fm]}\\\vspace{-3mm}
{BETA2B} & {-1}   & {deformation parameter $\beta_{2}$, nucleus $B$}\\\vspace{-3mm}
{BETA4B} & {-1}   & {deformation parameter $\beta_{4}$, nucleus $B$}\\\vspace{-3mm}
{ROTB\_THETA} & {-1} & {rotation parameter (angle $\theta$), -1 - random rotation, nucleus $B$} \\\vspace{-3mm} 
{ROTB\_PHI} & {-1} & {rotation parameter (angle $\phi$), -1 - random rotation, nucleus $B$} \\\vspace{-3mm} 
{SNN}    & {-1}  &  {NN (parton-parton) ``wounding'' cross section [mb]}\\\vspace{-3mm}
{SBIN}   & {-1}  &  {NN (parton-parton) binary cross section [mb]}\\\vspace{-3mm}
{ALPHA}  & {0.12}  &  {mixed model parameter, 0 - wounded, 1 - binary, 0.12 - LHC@2.76~TeV/nucleon}\\\vspace{-3mm}
{CD}     & {0.9}   &  {closest allowed distance between centers of nucleons [fm]}\\\vspace{-3mm}
{ISEED}  & {0}     &  {seed for the random number generator, if 0 a random seed is generated} \\\vspace{-3mm}
{BTOT}   &  &  range parameter for histograms [fm], default= max(RWSA,RWSB)+AWSA+AWSB\\\vspace{-3mm}
{SHIFT}  & {1}     &  {1 - shift the coordinates of the fireball to the c.m. in the fixed-axes case, 0 - do not shift} \\\vspace{-3mm} 

{DW}     & {0.}     &  {dispersion of the location of the source for wounded objectss [fm]}\\\vspace{-3mm}
{DBIN}   & {0.}     &  {dispersion of the location of the source for binary collisions [fm]}\\\vspace{-3mm}
{WFA}    & {0}     &  {the w parameter of the Fermi distribution, nucleus $A$}\\\vspace{-3mm}
{WFB}    & {0}     &  {the w parameter of the Fermi distribution, nucleus $B$}\\\vspace{-3mm}

{FBIN}   & {72}    &  {number of bins for histograms in the azimuthal angle} \\\vspace{-3mm}
{DOBIN}  & {0}     &  {1 - compute the binary collisions also for the case ALPHA=0, 0 - do not} \\\vspace{-3mm}
{GA}     & {1.0}  &  {central value of the Gaussian wounding profile}\\\vspace{-3mm}
{PP}  & {-1}    & {power of the transverse radius in the Fourier moments} \\\vspace{-3mm}
{RO}  & {0}    & {rank of the rotation axes (0 - rotation rank = rank of the Fourier moment)} \\\vspace{-3mm} 
{RCHA}   & {5.66}  & {harmonic oscillator shell model density mean squared charge radii of nucleus $A$ (${}^{12}$C-nucleus)}\\\vspace{-3mm}
{RCHB}   & {5.66}  & {harmonic oscillator shell model density mean squared charge radii of nucleus $B$ (${}^{12}$C-nucleus)} \\\vspace{-3mm} 
{RCHP}   & {0.7714} & {harmonic oscillator shell model density mean squared charge radii of  proton} \\\vspace{-3mm} 
{OMEGA}  & {-1}   & {relative variance of cross-section fluctuations for the Gamma wounding profile} \\\vspace{-3mm} 
{GAMA}   & {-1}     & {central value of the Gamma wounding profile}\\\vspace{-3mm}
{SCALEA} & {1.}   &  {scale parameter for the size of the nucleus (cluster version) [fm]} \\\vspace{-3mm} 
{SIGMAA} & {1.}   &  {standard deviation of x, y, z coordinates of nucleons in the alpha cluster [fm] } \\\vspace{-3mm} 
{SIGMABISA} & {1.}  & {standard deviation of x, y, z coordinates of nucleons in the $^3$He cluster or nucleon no. 9 [fm]} \\\vspace{-3mm} 
{SCALEB} & {1.}   & {scale parameter for the size of the nucleus (cluster version) [fm]} \\\vspace{-3mm} 
{SIGMAB} & {1.}   & {standard deviation of x, y, z coordinates of nucleons in the alpha cluster  [fm]} \\\vspace{-3mm} 
{SIGMABISB} & {1.} & {standard deviation of x, y, z coordinates of nucleons in the $^3$He cluster or nucleon no. 9 [fm]} \\\vspace{-3mm} 
{QSCALE} & {-1} & {scale parameter in the parton distribution function [fm]} \\\vspace{-3mm} 
{DS}     & {-1}    & {source smearing parameter} \\
\\
             
\hline

\end{longtable}
}

\bibliography{hydr,hydr1,hydr0,hydro,literHBT,mrlit,liter}

\begin{thebibliography}{100}

\bibitem{Broniowski:2007nz}
W. Broniowski, M. Rybczy\'nski and P. Bo\.zek,
\newblock Comput. Phys. Commun. 180 (2009) 69, 0710.5731.

\bibitem{Rybczynski:2013yba}
M. Rybczy\'nski et~al.,
\newblock Comput. Phys. Commun. 185 (2014) 1759, 1310.5475.

\bibitem{glauber1959high}
R. Glauber,
\newblock {High energy collision theory,~}volume 1 in Lectures in theoretical
  physics (Interscience, NewYork, 1959).

\bibitem{Czyz:1969jg}
W. Czy\.z and L.C. Maximon,
\newblock Annals Phys. 52 (1969) 59.

\bibitem{Miller:2007ri}
M.L. Miller et~al.,
\newblock Ann. Rev. Nucl. Part. Sci. 57 (2007) 205, nucl-ex/0701025.

\bibitem{Alver:2008aq}
B. Alver et~al.,
\newblock (2008), 0805.4411.

\bibitem{Loizides:2014vua}
C. Loizides, J. Nagle and P. Steinberg,
\newblock SoftwareX 1-2 (2015) 13, 1408.2549.

\bibitem{Loizides:2017ack}
C. Loizides, J. Kamin and D. d'Enterria,
\newblock Phys. Rev. C97 (2018) 054910, 1710.07098,
\newblock [erratum: Phys. Rev.C99,no.1,019901(2019)].

\bibitem{Moreland:2014oya}
J.S. Moreland, J.E. Bernhard and S.A. Bass,
\newblock Phys. Rev. C92 (2015) 011901, 1412.4708.

\bibitem{Wang:1991hta}
X.N. Wang and M. Gyulassy,
\newblock Phys. Rev. D44 (1991) 3501.

\bibitem{Lin:2004en}
Z.W. Lin et~al.,
\newblock Phys. Rev. C72 (2005) 064901, nucl-th/0411110.

\bibitem{Bass:1998ca}
S.A. Bass et~al.,
\newblock Prog. Part. Nucl. Phys. 41 (1998) 255, nucl-th/9803035.

\bibitem{Werner:2010aa}
K. Werner et~al.,
\newblock Phys. Rev. C82 (2010) 044904, 1004.0805.

\bibitem{Bialas:1977en}
A. Bia\l{}as, W. Czy\.z and W. Furma\'nski,
\newblock Acta Phys. Polon. B8 (1977) 585.

\bibitem{Bialas:1977xp}
A. Bia\l{}as et~al.,
\newblock Acta Phys. Polon. B8 (1977) 855.

\bibitem{Anisovich:1977av}
V.V. Anisovich, {\relax Yu}.M. Shabelski and V.M. Shekhter,
\newblock Nucl. Phys. B133 (1978) 477.

\bibitem{Bialas:1978ze}
A. Bia\l{}as and W. Czy\.z,
\newblock Acta Phys. Polon. B10 (1979) 831.

\bibitem{Eremin:2003qn}
S. Eremin and S. Voloshin,
\newblock Phys. Rev. C67 (2003) 064905, nucl-th/0302071.

\bibitem{KumarNetrakanti:2004ym}
P. Kumar~Netrakanti and B. Mohanty,
\newblock Phys. Rev. C70 (2004) 027901, nucl-ex/0401036.

\bibitem{Adler:2013aqf}
PHENIX, S.S. Adler et~al.,
\newblock Phys. Rev. C89 (2014) 044905, 1312.6676.

\bibitem{Adare:2015bua}
PHENIX, A. Adare et~al.,
\newblock Phys. Rev. C93 (2016) 024901, 1509.06727.

\bibitem{Lacey:2016hqy}
R.A. Lacey et~al.,
\newblock Universe 4 (2018) 22, 1601.06001.

\bibitem{Bozek:2016kpf}
P. Bo\.zek, W. Broniowski and M. Rybczy\'nski,
\newblock Phys. Rev. C94 (2016) 014902, 1604.07697.

\bibitem{Zheng:2016nxx}
L. Zheng and Z. Yin,
\newblock Eur. Phys. J. A52 (2016) 45, 1603.02515.

\bibitem{Mitchell:2016jio}
J.T. Mitchell et~al.,
\newblock Phys. Rev. C93 (2016) 054910, 1603.08836.

\bibitem{Loizides:2016djv}
C. Loizides,
\newblock Phys. Rev. C94 (2016) 024914, 1603.07375.

\bibitem{Bialas:1976ed}
A. Bia\l{}as, M. B\l{}eszy\'nski and W. Czy\.z,
\newblock Nucl. Phys. B111 (1976) 461.

\bibitem{Bialas:2008zza}
A. Bia\l{}as,
\newblock J. Phys. G35 (2008) 044053.

\bibitem{Kharzeev:2000ph}
D. Kharzeev and M. Nardi,
\newblock Phys. Lett. B507 (2001) 121, nucl-th/0012025.

\bibitem{Back:2001xy}
PHOBOS, B.B. Back et~al.,
\newblock Phys. Rev. C65 (2002) 031901, nucl-ex/0105011.

\bibitem{Ollitrault:1992bk}
J.Y. Ollitrault,
\newblock Phys. Rev. D46 (1992) 229.

\bibitem{Aguiar:2001ac}
C.E. Aguiar et~al.,
\newblock Nucl. Phys. A698 (2002) 639, hep-ph/0106266.

\bibitem{Miller:2003kd}
M. Miller and R. Snellings,
\newblock (2003), nucl-ex/0312008.

\bibitem{Manly:2005zy}
PHOBOS, S. Manly et~al.,
\newblock Nucl. Phys. A774 (2006) 523, nucl-ex/0510031.

\bibitem{Andrade:2006yh}
R. Andrade et~al.,
\newblock Phys. Rev. Lett. 97 (2006) 202302, nucl-th/0608067.

\bibitem{Voloshin:2006gz}
S.A. Voloshin,
\newblock (2006), nucl-th/0606022.

\bibitem{Alver:2006wh}
PHOBOS Collaboration, B. Alver et~al.,
\newblock Phys. Rev. Lett. 98 (2007) 242302, nucl-ex/0610037.

\bibitem{Drescher:2006ca}
H.J. Drescher and Y. Nara,
\newblock Phys. Rev. C75 (2007) 034905, nucl-th/0611017.

\bibitem{Broniowski:2007ft}
W. Broniowski, P. Bo\.zek and M. Rybczy\'nski,
\newblock Phys. Rev. C76 (2007) 054905, 0706.4266.

\bibitem{Voloshin:2007pc}
S.A. Voloshin et~al.,
\newblock Phys. Lett. B659 (2008) 537, 0708.0800.

\bibitem{Andrade:2008fa}
R.P.G. Andrade et~al.,
\newblock Acta Phys. Polon. B40 (2009) 993, 0812.4143.

\bibitem{Alver:2008zza}
PHOBOS, B. Alver et~al.,
\newblock Phys. Rev. C77 (2008) 014906, 0711.3724.

\bibitem{Broniowski:2009fm}
W. Broniowski, M. Chojnacki and L. Obara,
\newblock Phys. Rev. C80 (2009) 051902, 0907.3216.

\bibitem{Andrade:2009em}
R. Andrade et~al.,
\newblock J.Phys. G37 (2010) 094043, 0912.0703.

\bibitem{Hama:2009pk}
Y. Hama et~al.,
\newblock Acta Phys. Polon. B40 (2009) 931, 0901.2849.

\bibitem{Hirano:2009ah}
T. Hirano and Y. Nara,
\newblock Phys.Rev. C79 (2009) 064904, 0904.4080.

\bibitem{Alver:2010gr}
B. Alver and G. Roland,
\newblock Phys. Rev. C81 (2010) 054905, 1003.0194.

\bibitem{Alver:2010dn}
B.H. Alver et~al.,
\newblock Phys. Rev. C82 (2010) 034913, 1007.5469.

\bibitem{Staig:2010pn}
P. Staig and E. Shuryak,
\newblock Phys.Rev. C84 (2011) 034908, 1008.3139.

\bibitem{Teaney:2010vd}
D. Teaney and L. Yan,
\newblock Phys. Rev. C83 (2011) 064904, 1010.1876.

\bibitem{Qin:2010pf}
G.Y. Qin et~al.,
\newblock Phys.Rev. C82 (2010) 064903, 1009.1847.

\bibitem{Nagle:2010zk}
J.L. Nagle and M.P. McCumber,
\newblock Phys.Rev. C83 (2011) 044908, 1011.1853.

\bibitem{Xu:2010du}
J. Xu and C.M. Ko,
\newblock Phys.Rev. C83 (2011) 021903, 1011.3750.

\bibitem{Lacey:2010av}
R.A. Lacey et~al.,
\newblock Phys.Rev. C84 (2011) 027901, 1011.3535.

\bibitem{Adare:2010ux}
PHENIX Collaboration, A. Adare et~al.,
\newblock Phys. Rev. Lett. 105 (2010) 062301, 1003.5586.

\bibitem{Qin:2011uw}
G.Y. Qin and B. Muller,
\newblock Phys.Rev. C85 (2012) 061901, 1109.5961.

\bibitem{Bhalerao:2011bp}
R.S. Bhalerao, M. Luzum and J.Y. Ollitrault,
\newblock Phys. Rev. C84 (2011) 054901, 1107.5485.

\bibitem{Bhalerao:2011yg}
R.S. Bhalerao, M. Luzum and J.Y. Ollitrault,
\newblock Phys. Rev. C84 (2011) 034910, 1104.4740.

\bibitem{Gardim:2011xv}
F.G. Gardim et~al.,
\newblock Phys. Rev. C85 (2012) 024908, 1111.6538.

\bibitem{Qiu:2011hf}
Z. Qiu, C. Shen and U. Heinz,
\newblock Phys. Lett. B707 (2012) 151, 1110.3033.

\bibitem{Qiu:2011iv}
Z. Qiu and U.W. Heinz,
\newblock Phys.Rev. C84 (2011) 024911, 1104.0650.

\bibitem{Xu:2011jm}
J. Xu and C.M. Ko,
\newblock Phys.Rev. C84 (2011) 044907, 1108.0717.

\bibitem{Teaney:2012ke}
D. Teaney and L. Yan,
\newblock Phys. Rev. C86 (2012) 044908, 1206.1905.

\bibitem{Jia:2012ma}
J. Jia and S. Mohapatra,
\newblock (2012), 1203.5095.

\bibitem{Hirano:2012kj}
T. Hirano et~al.,
\newblock Prog. Part. Nucl. Phys. 70 (2013) 108, 1204.5814.

\bibitem{Broniowski:2013dia}
W. Broniowski and E.R. Arriola,
\newblock Phys. Rev. Lett. 112 (2014) 112501, 1312.0289.

\bibitem{Bozek:2014cva}
P. Bo{\.z}ek et~al.,
\newblock Phys.Rev. C90 (2014) 064902, 1410.7434.

\bibitem{Zhang:2017xda}
S. Zhang et~al.,
\newblock Phys. Rev. C95 (2017) 064904, 1702.02507.

\bibitem{Rybczynski:2017nrx}
M. Rybczy\'nski, M. Piotrowska and W. Broniowski,
\newblock Phys. Rev. C97 (2018) 034912, 1711.00438.

\bibitem{Lim:2018huo}
S.H. Lim et~al.,
\newblock Phys. Rev. C99 (2019) 044904, 1812.08096.

\bibitem{Adare:2015ctn}
PHENIX, A. Adare et~al.,
\newblock (2015), 1507.06273.

\bibitem{Wiringa:2013ala}
R.B. Wiringa et~al.,
\newblock Phys. Rev. C89 (2014) 024305, 1309.3794.

\bibitem{Lonardoni:2017egu}
D. Lonardoni et~al.,
\newblock Phys. Rev. C96 (2017) 024326, 1705.04337.

\bibitem{Patrignani:2016xqp}
Particle Data Group, C. Patrignani et~al.,
\newblock Chin. Phys. C40 (2016) 100001.

\bibitem{Werner:2009fa}
K. Werner et~al.,
\newblock J. Phys. G36 (2009) 064030, 0907.5529.

\bibitem{Petersen:2010cw}
H. Petersen et~al.,
\newblock Phys. Rev. C82 (2010) 041901, 1008.0625.

\bibitem{Holopainen:2010gz}
H. Holopainen, H. Niemi and K.J. Eskola,
\newblock Phys. Rev. C83 (2011) 034901, 1007.0368.

\bibitem{Bozek:2011if}
P. Bo\.zek,
\newblock Phys. Rev. C85 (2012) 014911, 1112.0915.

\bibitem{Schenke:2010rr}
B. Schenke, S. Jeon and C. Gale,
\newblock Phys. Rev. Lett. 106 (2011) 042301, 1009.3244.

\bibitem{Qiu:2011fi}
Z. Qiu and U.W. Heinz,
\newblock AIP Conf.Proc. 1441 (2012) 774, 1108.1714.

\bibitem{Chaudhuri:2011pa}
A. Chaudhuri,
\newblock Phys. Lett. B713 (2012) 91, 1112.1166.

\bibitem{na61url:2013}
\url{https://na61.web.cern.ch/na61/xc/index.html},
\newblock {NA61/SHINE-Collaboration}.

\bibitem{Filip:2007tj}
P. Filip,
\newblock Phys. Atom. Nucl. 71 (2008) 1609, 0712.0088.

\bibitem{Filip:2009zz}
P. Filip et~al.,
\newblock Phys. Rev. C80 (2009) 054903.

\bibitem{Filip:2010zz}
P. Filip,
\newblock Nucl. Phys. Proc. Suppl. 198 (2010) 46.

\bibitem{Alvioli:2009ab}
M. Alvioli, H.J. Drescher and M. Strikman,
\newblock Phys. Lett. B680 (2009) 225, 0905.2670.

\bibitem{Broniowski:2010jd}
W. Broniowski and M. Rybczy\'nski,
\newblock Phys. Rev. C81 (2010) 064909, 1003.1088.

\bibitem{Hohne:2006ks}
C. Hohne, F. Puhlhofer and R. Stock,
\newblock Phys. Lett. B640 (2006) 96, hep-ph/0507276.

\bibitem{Becattini:2008ya}
F. Becattini and J. Manninen,
\newblock Phys. Lett. B673 (2009) 19, 0811.3766.

\bibitem{Bozek:2005eu}
P. Bo\.zek,
\newblock Acta Phys. Polon. B36 (2005) 3071, nucl-th/0506037.

\bibitem{Werner:2007bf}
K. Werner,
\newblock Phys. Rev. Lett. 98 (2007) 152301, 0704.1270.

\bibitem{doxygen:2013}
\url{http://www.stack.nl/~dimitri/doxygen/}.

\bibitem{Antchev:2013paa}
TOTEM, G. Antchev et~al.,
\newblock Phys. Rev. Lett. 111 (2013) 012001.

\bibitem{Antchev:2015zza}
TOTEM, G. Antchev et~al.,
\newblock Nucl. Phys. B899 (2015) 527, 1503.08111.

\bibitem{Aad:2014dca}
ATLAS, G. Aad et~al.,
\newblock Nucl. Phys. B889 (2014) 486, 1408.5778.

\bibitem{Collaboration:2012wt}
Pierre Auger, P. Abreu et~al.,
\newblock Phys. Rev. Lett. 109 (2012) 062002, 1208.1520.

\bibitem{Rybczynski:2013mla}
M. Rybczy\'nski and Z. W\l{}odarczyk,
\newblock J.Phys. G41 (2013) 015106, 1307.0636.

\bibitem{Rybczynski:2011wv}
M. Rybczynski and W. Broniowski,
\newblock Phys. Rev. C84 (2011) 064913, 1110.2609.

\bibitem{Heinz:2004ir}
U.W. Heinz and A. Kuhlman,
\newblock Phys. Rev. Lett. 94 (2005) 132301, nucl-th/0411054.

\bibitem{Rybczynski:2012av}
M. Rybczy\'nski, W. Broniowski and G. Stefanek,
\newblock Phys. Rev. C87 (2013) 044908, 1211.2537.

\bibitem{Zhang:2018zzu}
S. Zhang et~al.,
\newblock Eur. Phys. J. A54 (2018) 161, 1808.10265.

\bibitem{root1}
\url{https://root.cern.ch/}.

\bibitem{Carlson:1997qn}
J. Carlson and R. Schiavilla,
\newblock Rev. Mod. Phys. 70 (1998) 743.

\end{thebibliography}

\end{document}